# The Effect of a Surface Tension Gradient on the Slip Flow along a Superhydrophobic Air-Water Interface

Dong Song[a,b,c], Baowei Song[a,*], Haibao Hu[a]*, Xiaosong Du[d], Peng Du[e], Chang-Hwan Choi[c] and Jonathan P. Rothstein[b]*

[a] *School of Marine Science and Technology, Northwestern Polytechnical University, 127 Youyi Xilu, Xi'an, 710072, Shaanxi, P. R. China*

[b] *Department of Mechanical and Industrial Engineering, University of Massachusetts Amherst, 160 Governors Drive, Amherst, MA 01003, USA*

[c] *Department of Mechanical Engineering, Stevens Institute of Technology, Hoboken, New Jersey 07030, United States*

[d] *Microproducts Breakthrough Institute, Corvallis, Oregon, 97330, USA*

[e] *Sorbonne Universities, University of Technology of Compiegne, Lab. Roberval, UMR 7337 CNRS Centre de Recherches Royallieu BP 20529, 60206 Compiegne Cedex, France*

\* To whom correspondence should be addressed:

hanghai@nwpu.edu.cn (Baowei Song), huhaibao@nwpu.edu.cn (Haibao Hu),

rothstein@ecs.umass.edu (Jonathan P. Rothstein).

# ABSTRACT

Superhydrophobic surfaces have been shown to produce significant drag reduction in both laminar and turbulent flows by introducing an apparent slip velocity along an air-water interface trapped within the surface roughness. In the experiments presented within this study, we demonstrate the existence of a surface tension gradient associated with the resultant Marangoni flow along an air-water interface that causes the slip velocity and slip length to be significantly reduced. In this study, the slip velocity along a millimeter-sized air-water interface was investigated experimentally. This large-scale air-water interface facilitated detailed investigation of the interfacial velocity profiles as the flow rate, interfacial curvature and interface geometry were varied. For the air-water interfaces supported above continuous grooves (concentric rings within a torsional shear flow) where no surface tension gradient exists, slip velocity as high as 30% of the bulk velocity was observed. However, for the air-water interfaces supported above discontinuous grooves (rectangular channels in a Poiseuille flow), the presence of a surface tension gradient reduced the slip velocity and in some cases resulted in an interfacial velocity that was opposite to the main flow direction. The curvature of the air-water interface in the spanwise direction was found to dictate the details of the interfacial flow profile with reverse flow in the center of the interface for concave surfaces and along the outside of the interface for convex surfaces. The deflection of the air-water interface was also found to greatly affect the magnitude of the slip. Numerical simulations imposed with a relatively small surface tension gradient along air-water interface were able to predict both the reduced slip velocity and back flow along the air-water interface.

## I. INTRODUCTION

Superhydrophobic surfaces, composed of low surface energy materials and microstructures, can trap air within the microstructures. These stable air pockets, sandwiched between the fluid and solid surface asperities, promise to change the classical no-slip boundary to a slip boundary at the air-water interfaces of the superhydrophobic surface under water. In the past score years, masses of theoretical analysis, numerical simulations and experimental investigations have indicated the ability of superhydrophobic surfaces to reduce the frictional drag in laminar flows in microfluidics and rheometers, as well as turbulent flows in pipes, channels or over plates in a towing tank [1-10]. A series of measurements have been performed to explore the details of the boundary layer near the superhydrophobic surfaces and the slip velocity at the micro air pockets within the surface asperities has been successfully measured using particle image velocimetry (micro-PIV). Significant slip velocity has been directly observed at the superhydrophobic surfaces and the slippage will reduce the shear stress, turbulence production, or strength and patterns of turbulent vortex [1,11-18].

It should be noted, however, that the expected drag reduction or slip velocity at the superhydrophobic surfaces was not obtained according to a growing body of work. It was postulated that the air-water interface cannot be treated as shear free in some cases, but that a finite shear stress can exist at the air-water interface especially in the presence of surfactants, particles or other surface-active agents [19-26]. In an open channel, Yang et al. [24] found that the no-slip boundary condition at the air-water interface was in better agreement with the experimental data than the shear-free boundary condition. Bolognesi et al. [19,25] measured the velocity profiles and shape of the air-water interfaces on the superhydrophobic surfaces using a micro-PIV system and demonstrated that the shear stress was non-zero along the air-water interface. They postulated that the non-zero shear stress was introduced by a Marangoni flow resulting from the presence of surfactants in the water. However the presence and spatial distribution of surfactants along the air-water interface within a microfluidic device are extremely difficult to directly observe. Very recently, Peaudecerf et al. [27] showed that even a very small amount of surfactant could yield a no-slip boundary condition at a flat air-water interface. Similar results have been confirmed by detecting the viscoelastic response to a vibrating interface using AFM [26]. However, in the work mentioned above, the investigated air-water interfaces at the superhydrophobic surfaces are in micro scale and the effect of their deformation is not examined. In this work, the impact of the streamwise surface tension gradient at a curved air-water interface was investigated by considering both the deflection and shape of the interface. It shows that the slip velocity at the air-water interface in the rectangular channel is very small, while the slip velocity at a similar-sized circular air-water interface is dramatically large in a torsional shear flow.

## II. EXPERIMENTAL SETUP

### 2.1 Channel fabrication

In this work, a modified confocal micro-PIV system was developed to measure the slip velocity along a single air-water interface. A millimeter wide groove was fabricated into either the one wall (side or bottom) of a hydrophobic channel where it supported a stable air-water interface. The hydrophobic channels were made of transparent PDMS (Sylgard 184 Dow Corning) which had an advancing contact angle around 118°. For the first channel, shown schematically in Fig. 1a, an air bubble would be trapped in the horizontal groove (1.0 mm wide, 5.0 mm long and 5.0 mm deep) on the side wall of the channel. In this orientation, the velocity profile in the cross section normal to the air-water interface could be measured. To make the groove on the side wall, a 7.0 mm × 5.0 mm × 1.0 mm tiny glass block was placed horizontally on a glass slide with a piece of 1.0 mm thick solid PDMS beneath. The liquid PDMS was poured onto the glass slide with a 3.0 mm thick fence. After the PDMS was cured in an oven at 60 ℃ for 3 hours, the glass block was removed leaving a groove at the side, and the channel itself was created by cutting out a rectangular strip (5.5 mm×58 mm) along the center line. In the end, the channel was sealed by covering another piece of PDMS on the top of the channel.

For the second channel, shown schematically in Fig. 1b, a vertical groove (1.0 mm wide, 15.0 mm long and 2.0 mm deep) was fabricated in the bottom wall of the channel. PDMS 2.0 mm thick was initially cured on a glass slide as the bottom layer. The groove was created at the bottom layer by cutting a 1.0 mm wide and 15.0 mm long strip along the center line. A middle layer, with a 5 mm × 56 mm groove (as the channel), and a top layer (a piece of 1.1 mm thick PDMS) were sandwiched together on the bottom layer. The channel was sealed by pouring liquid PDMS between the gaps of each layer and cured in the oven at 60 ℃ for around 3 hours. In this orientation, the velocity profile across the whole air-water interface could be measured. In the discussion that follows, the coordinates are as marked in Fig. 1: $x$ is always in the streamwise direction and $z$ is always in the direction opposite to that of gravity. Due to the changing orientation of the flow cells, this means that $y$ is either normal to the air-water interfaces as it is for the case where the interface is on the side wall (Fig. 1a) or in the shearing direction tangent to the air-water interface as it is for the case where the interface is along the bottom wall of the channel (Fig. 1b). The bulk velocity, $U_\infty$, was measured by the volumetric flow rate through outlet of the channel. In all of the tests, the Reynolds number, $Re = \rho U_\infty H / \mu$, was controlled to be smaller than 2 to make the flow laminar, where $\rho$, $H$, and $\mu$, are the density of water, channel height and dynamic viscosity of water, respectively. And the capillary number, $Ca = \mu U_\infty / \gamma$, with $\gamma$ being the air-water surface tension, was controlled between $0.43 \times 10^{-5}$ to $1.3 \times 10^{-5}$, under which condition the flow was dominated by both the viscous force and surface tension, and the deflection of the air-water interface on top of the groove could be easily manipulated at the same time.

The slip velocity at the continuous air-water interface within a torsional shear flow was also measured in a circular channel as shown in Fig. 1c. The circular channel was made of polymethyl methacrylate (PMMA). The bottom plate was fixed while the top plate was rotated by a driving wheel. The diameters of both the bottom and top plate are 40 mm. An annular groove with a radius of the center-line, $r_c$ = 8 mm and a width, $W_g$, in the range between 0.94 mm and 1.83 mm, was fabricated at the bottom of the channel. The gap between the top and bottom plates was adjusted between 1.0 mm < $H$ < 4.0 mm. The rotating velocity of the top plate was adjusted between 0.036 rad/s < $\Omega$ < 0.210 rad/s corresponding to the average tangential velocity of the top plate at the groove area varying between 0.29 mm/s < $U_\infty = \Omega r_c$ < 1.67 mm/s with corresponding $Re$ in the range of 0.28 and 6.72 and $Ca$ in the range of $0.39 \times 10^{-5}$ and $2.3 \times 10^{-5}$. It should be noted that a commercial superhydrophobic coating, Ultra-Ever Dry (Ultratech International, Inc.) [28], was coated on the bottom surface to make the air-water interface durable enough to keep it from collapsing. A 2.0 mm-diameter stainless steel cylinder was inserted into the center of both the top and bottom surface with a 3 mm outer-diameter ring acting as a spacer between the top and bottom surfaces.

## 2.2 Air-water interface manipulation

In order to properly measure the slip velocity along the air-water interface, two key concerns needed to be addressed. First, the size of the air-water interface was purposely made quite large so that the details of the slip velocity and the surface curvature could be easily observed. At this scale, the air-water interface can collapse easily even under a small static loading pressure and can become unstable in dynamic flows with modest pressure fluctuations. This is one of the major reasons why the superhydrophobic surfaces with larger-scale microstructures have not been applied on engineering applications even though drag reduction is known to scale with feature size [29-33]. In order to maintain the air-water interface, during filling, the channels were oriented such that the millimeter-sized groove on the top while the fluid was slowly injected. Once the channel was fully filled with water, the channel was rotated to the desired orientation. Using this procedure, the air-water interface was found to be sufficiently durable to support the fluid both under static and flow conditions. However, to further insure the stability of the interface and to decrease the fluctuation of the pressure as much as possible, the flow was driven by gravity instead of a mechanical pump. To the best of our knowledge, this is the first time that a direct measurement of the slip velocity along such a large scale (millimeter-sized) air-water interface has been performed within a closed channel.

Secondly, due to the size of the air-water interface in these experiments, even small differences in pressure between the water and air can cause the interface to deform making it a challenge to keep it flat. This, however, is also an opportunity as the deflection of the air-water interface can be precisely manipulated through changes to the volume of the air in the groove induced by altering the static pressure of the water within the flow field. According to the ideal gas law or Boyle's law, the pressure and volume are inversely proportional at a constant temperature, $P \sim 1/V$. The Laplace pressure generated by the deflection of the air-water interface on millimeter scale is small enough to be neglected. Thus, by lowering or increasing the pressure, the volume of the air in the groove can be precisely controlled and the convex or concave curvature of the air-water interface can be accurately manipulated simultaneously. The static pressure within the channel and the flow velocity were adjusted by altering the height of the outlet and inlet terminals of the gravity fed flow.

To calculate the deflection of the air-water interface, the first step is to detect the critical pressure under which the interface is flat. As shown in Fig. 2, the laser initially lights the flow field downward in a slant angle perpendicular to the groove in the $y$-$z$ plane from the left side of the interface. When the air-water interface deflects upward ($\Delta d > 0$, convex), strong reflection appears at the area near the three phase contact line on the left side, as shown in Fig. 2b. However, the strong reflection will disappear and switch to the other side when the air-water interface deflects downward ($\Delta d < 0$, concave). The air-water interface is regarded as flat at the critical pressure under which the reflection position starts to transfer. The measured critical gage pressure is 3.0 kPa in the channel shown in Fig. 1b at 1.06 mm/s. It should be noted that the laser was changed to be parallel to the interface in the $x$-$z$ plane when measuring the slip velocity, so that there is no reflection area during measuring.

The value of the interface deflection, $\Delta d$, was not able to be measured directly in our experiments. However the interface deflection could be calculated from the volume change when altering the pressure by assuming the interface being circular arc. As shown in Fig. 2a, the initial volume of the air in the groove is $V_0 = W_g D_g L_g$ when the interface was flat at the absolute pressure $P_0 = 104.3$ kPa, where $W_g$, $D_g$, $L_g$ were the dimensions of the groove. Under pressure of $P_1$, the volume of the air in the groove changes to $V_1 = V_0 + (\theta R^2 - 0.5W(R - \Delta d)) \cdot L_g$, where the definitions of $R$ and $\theta$ are shown in Fig. 2a. Assuming the interface being circular arc, we can get $R^2 = (0.5W_g)^2 + (R - \Delta d)^2$ and $\theta = \arccos((R - \Delta d)/R)$. According to the ideal gas law $P_1 \cdot V_1 = P_0 \cdot V_0$, the deflection can be calculated directly. In this work, the gage pressure ($\Delta P$) is varied from -5.0 kPa to 7.0 kPa and the interface deflection ($\Delta d$) changes between 124 μm and -56 μm.

## 2.3 Slip velocity measurements

A modified confocal micro-PIV system was employed to measure the velocity in the field of view with millimeter size. The main difference between the standard micro-PIV system [18] and ours is that the microscope was replaced by a lens with a long working distance (25 mm) and a zoom-in capability of 300X. Unlike the standard micro-PIV system with a depth of field that can be less than 1 μm depending on magnification [34,35], the depth of field of our lens is approximately 150 μm. This large depth of field enables us to observe the whole air-water interface in a single image even when the interface is deeply curved. To illuminate the flow, the fluorescent polystyrene particles (MV-F07, Microvec Pte Ltd. China) with an average diameter of 7 μm and a density of 1.05 g/cm$^3$ were dispersed in deionized water. The excitation and emission wavelength of the fluorescent particles are 532 nm and 590 nm, respectively. The laser beam (~0.5 W) excited the flow field within a diameter of 2.0 mm at the air-water interface and the background laser light was filtered by an orange optical filter fixed on the microscope lens. Only the particles within the focal layer could be captured, while the particles out of focus would appear as either larger, discrete points or a background glow in the captured image [17]. The images were captured by a high speed camera (MotionXtra NX-4, IDT Corporation ) at 800×600 pixels with a spatial resolution of approximately 5 μm. The frequency of the image acquisition was based on the type of the channel which would be described in the following. The image sequences were then processed using open software PIVlab [36,37]. Each of the experiments was repeated at least three times to ensure repeatability and statistical analysis.

It is worth mentioning that the particles within the focus layer while not locating at the interface will affect our slip velocity measurements, which may result in a larger slip velocity than the real value. However, this error can be negligible based on two reasons. First of all, the concentration of the particles sticking at the interface is much larger than the ones in the fluid field as shown in Fig. 3a and the videos in the supporting information [38]. The other reason, as will be mentioned in the following, is that the slip velocity at the air-water interface shown in Fig. 1b is much smaller than the bulk velocity. As a result, the particles out of the interface move much faster than the ones sticking at the interface. We get rid of these particles by setting a maximum threshold value ($u_{s,\text{thr}} = 0.10 U_\infty$, for the channel shown in Fig. 1b), over which the velocity is not included in our results.

## III. SIMULATION

Numerical simulations were performed to compare with experimental results by means of the commercial package Fluent™ (Fluent Inc., New Hampshire, USA) using the steady laminar model, similar with the one made by Ou et al. [17]. The flow field was meshed via Gambit$^{\text{TM}}$ in structured grids. The dimensions of the channel and the interface are same as the experimental channels. For the simulations related to the rectangular channel flow shown in Figs. 1a and 1b, the air-water interface was treated as rigid with an arc surface in the middle part. At the head and tail area of the interface, the interface transforms from flat (solid-air-fluid edge) to arc (middle part) based on the spline function of Gambit and the transition region is 1.5 mm long. The inlet was simplified to be a uniform flow with value of $U_\infty$. The number of meshes at the interface area is $15 \times 20 \times 30$ for the model shown in Fig. 1a, and $15 \times 50 \times 30$ for the model shown in Fig. 1b. For the simulation related to the flow shown in Fig. 1c, the dimension is same with the experimental channel. The air-water interface was flat and the top surface was set to be a rotational wall.

Three kinds of boundary conditions were applied at the air-water interface. The first one was the standard no-slip condition that was the same as a common solid surface without any slippage. The second one was the shear-free condition on a rigid surface, which had been widely used for the simulations of the flow on the superhydrophobic surfaces [17,39]. A large slip velocity is expected to exist at the interface with the shear-free boundary condition. However, as it will be shown in the next section, the slip velocity at the experimental air-water interface is much smaller than the numerical one at the interface with shear-free boundary condition. For the third boundary condition, the experimental slip velocity (slip-defined) or calculated shear stress (shear-defined) was specified at the air-water interface using a user-defined function (UDF) in Fluent[TM]. This boundary condition can be used to calculate the shear stress or the surface tension gradient at the air-water interface by combining experimentally measured slip velocity.

## IV. RESULTS AND DISCUSSION

The velocity profile normal to the air-water interface was measured in the channel shown in Fig. 1a. The air-water interface, defined by the groove in the side wall, was adjusted to deflect ~0.3 mm towards the flow field in order to capture the interface clearly. The focal plane was kept normal to the interface and locating at the position with the maximum deflection. The images were captured at 100 fps and the velocity correlations were time-averaged over a minimum of 500 frames. One of the captured images along with the calculated velocity vector field is shown in Fig. 3a. Note that the concentration of the tracer particles at the air-water interface was much higher than other areas (see Video1.mp4 in the Supporting Information [38]). Fig. 3b shows the velocity profile as a function of the point-plane distance from the air-water interface. The experimental slip velocity was found to be much smaller than expected from theory with a maximum slip velocity of only 23% $U_\infty$. This result was robust and appeared to be independent of the bulk velocity in the channel as shown by the filled symbols in Fig. 3b for bulk velocity in the range of 0.30 mm/s $< U_\infty <$ 0.56 mm/s. The experimental slip length within the measured plane was calculated to be $b_{\text{exp}} = 0.09 \pm 0.03$ mm for all cases measured. Here, the local slip length, $b$, is defined using Navier's slip boundary condition, $b = u_s / |\partial u/\partial y|_{y=0}$, where $u_s$ is the slip velocity and $|\partial u/\partial y|_{y=0}$ is the velocity gradient at the interface [40]. Note that this value is inconsistent with the theoretical work of former people. For a shear-free interface, $|\partial u/\partial y|_{y=0} = 0$, the slip length should be infinite. The channel flow shown in Fig. 3 was numerically simulated. In the simulation, the dimension of the channel and the air-water interface was same with the experiments. The air-water interface was set shear-free in this condition. As shown by the hollow symbols in Fig. 3b, the numerical slip velocity at the center of the shear-free interface reaches up to 119 % of the average bulk velocity and $|\partial u/\partial y| = 0$ at the interface. This is qualitatively different with the experimental measurements. In fact, as seen in Fig. 3b, the experimental measurements are much closer to the predictions of the numerical simulations performed with the no-slip boundary condition than the shear-free boundary condition.

This contradiction indicates that the shear-free assumption at the air-water interface is inappropriate and a shear stress, $\tau_{AW}$, exists at the air-water interface and that this shear stress is quite close to the shear stress generated by the velocity gradient above a solid wall without any slip. The origin of the shear stress may result from the presence of unknown surface-active agents which has created surface tension gradient, $\partial \gamma / \partial x$, at the air-water interface as suggested by Bolognesi et al. [19] and discussed recently by Schaffel et al. [41] and Peaudecerf et al. [27] for the interfaces contaminated by surfactants. The presence of a surface tension gradient along the air-water interface can cause a shear stress along the interface. This is known as the Marangoni effect and is usually caused by a flow-induced gradient in the concentration of a surface-active species at a fluid-fluid interface or a temperature gradient along the interface. Surface tension gradients can lead to interfacial flow in the direction of increasing surface tension. Often this interfacial flow is in the direction opposite to the direction of the bulk flow as is the case for an air bubble rising in water coated with surfactant. As the bubble rises, the surfactant is swept to the back of the bubble by the flow, creating a surface tension gradient and an interfacial flow that resists the rising motion of the bubble [42-44].

In order to better understand the interfacial flow, a number of additional investigations into the origins of the shear stress at the air-water interface were performed using the channel shown in Fig. 1b. This channel was developed so that the velocity profile across the entire air-water interface could be measured. In this channel, a 1.0 mm wide, 15.0 mm long and 2.0 mm deep groove was introduced in the bottom wall. Since the density of the particles is slightly higher than that of water, the particles were adsorbed slowly to the air-water interface. As can be seen in the images in Fig. 4 and Video2.mp4 of the Supporting Information [38], the particles at the air-water interface were clear and intensely fluorescent while the particles in the bulk flow at other areas within the focal layer were relatively sparse making it easy to clearly capture the flow at the air-water interface. One of the captured images alongside one of the time-averaged velocity vector fields at the interface is shown in Fig. 4a. In this experiment, the frame rate was only 3 fps and velocity correlations were averaged using at least 500 frames. The bulk velocity was kept constant at $U_\infty = 1.06$ mm/s. The gage pressure of the flow field changed in the range of 7.0 kPa and -5.0 kPa resulting in a deflection of the air-water interface between -56 μm $< \triangle d <$ 124 μm. Here, $\triangle d$ means the point-plane distance between the top of the groove and the center point of the interface as shown in Fig. 2a. The negative value of the interface deflection ($\triangle d$) indicates a concave interface curved down into the groove.

The velocity profiles along an air-water interface deflected by 92 μm into the bulk flow are shown in Fig. 4 for a channel with the groove at the bottom. In Fig. 4, we focus our measurements in the area just upstream where the groove ends and the air-water interface terminates at the head wall. At this point, we expect a buildup of particles or contaminant, which may cause a reduction in interfacial tension and the onset of a Marangoni flow. The slip velocity along the center line in the streamwise direction (line 1 marked in Fig. 4a) is shown in Fig. 4b. These measurements demonstrate that the effect of the head edge of the interface on the flow clearly hinders the slippage, resulting in a maximum of the slip velocity of only 2.5% $U_\infty$ at the head area. It is not unexpected that the slip velocity along the interface would decrease rapidly to zero as approaching the three phase contact line. What is unexpected, however, is a pair of vortices that formed in the corners just upstream of the head wall. This can clearly be seen by the velocity vector and streamlines in Fig. 4a (see also the video of this phenomenon found in Video2.mp4 of the Supporting Information [38]). The *x*-component of the velocity along the spanwise direction (line 2 marked in Fig. 4a) is shown in Fig. 4c. At the area close to the streamwise center line (0.2 < $y/W_g$ < 0.8), the slip is positive. However, at the area near the edge of the interface (0 < $y/W_g$ < 0.2 and 0.8 < $y/W_g$ < 1.0), the slip velocity is negative which means the slip flow was found to move in the direction opposite to the main flow. This reverse flow is clearly one of the main reasons for the dramatic reduction in the overall slip velocity and drag reduction observed for the contaminated superhydrophobic surfaces. It should be noted that according to the two phase flow simulations, where the air flow within the microstructures of the superhydrophobic surfaces was counted, apparent slip velocity was observed [41,45,46]. Therefore, the flow of the air trapped within the groove is not responsible for the reduction of the slip velocity, taking into consideration of the comparable sizes between the groove depth and the channel height in our experiments.

The reverse flow along the air-water interface can be found not only at the head area, but also the entire air-water interface. The slip velocity at the center of the air-water interface was shown in Fig. 5a (see also the video of this phenomenon in Video3.mp4 of the Supporting Information [38]). It is interesting that, the slip at the side area was also observed to remain opposite to the main flow direction even at the central area, far away from the head wall. Although the reversed slip flow was not observed either by Bolognesi et al. [19] or Schaffel et al. [41] in the presence of surfactants in a closed microchannel, this phenomenon has been noticed at an air-water interface exposed to the outside at an open channel [22,23]. Nevertheless the details of the reverse flow have not been further investigated up to now. As a matter of fact, the strength of the slip velocity, the size of the slip length and the details of the reverse flow along the air-water interface were all found to be quite sensitive to the curvature of the air-water interface as demonstrated by our experiments. When the pressure difference between the water and the air phase was increased from -5.0 kPa to 7.0 kPa, the interface deflection, $\triangle d$, was forced to decrease from 124 μm to -56 μm. As the interface changed from convex ($\triangle d > 0$) to concave ($\triangle d < 0$), the velocity profile along the air-water interface was found to transform dramatically as shown in Fig. 5b (also see Video4.mp4 of the Supporting Information [38]). Different from the slip velocity profile presented in Fig. 5a for a convex surface, the slip velocity profile for the concave interfaces was found to contain a reverse slip flow not along the sides of the air-water interface, but along the center line. In order to quantify these results further, the normalized slip velocity along the spanwise direction through the center of the interface is shown in Fig. 5c. It can be observed that the average slip velocity along the air-water interface increases as the deflection into the channel increases. Even so, the maximum of the slip velocity measured was only 3.0% of the average channel velocity, $U_\infty$, when the deflection was $\triangle d$ = 124 μm. When the air-water interface was deformed into the groove, $\triangle d < 0$, the maximum slip velocity was dramatically reduced becoming only 1.0% of average channel velocity when the deflection reached $\triangle d$ = -56 μm. For these concave interfaces, the reverse slip velocity was found to be comparable in magnitude to the positive slip velocity resulting in an average slip velocity that approaches zero or the no-slip boundary condition.

As width/height ratio of the channel shown in Fig. 1b is only 2.5, the flow field cannot be calculated using a simplified 2D model and it is impossible to give an analytical calculation of the slip length as well as the shear stress at the air-water interface based on our experimental data. However, the channel flow could be solved numerically by using the experimentally obtained slip velocity at the interface. The numerical simulations were performed using Fluent™ again. In this case, the geometry of the channel and the curvature of the interface were designed to be consistent with the experiments shown in Fig. 5c. Although the interfacial curvature was modeled, the air-water interface was modeled as a rigid surface with a finite slip velocity measured in the experiments using a user-defined function in Fluent (slip-defined boundary condition). A no-slip boundary condition was also used for comparison.

As shown in Fig. 6a, the average shear stress at the air-water interface with both slip-defined and no-slip boundary conditions was found to be directly proportional to the deflection. As deflection of the interface increases, the local height of the channel (distance between the top surface and interface) at the interface area decreases, resulting in the increase of both the local flow rate and the local shear stress. Because the shear stresses for no-slip and slip-defined conditions are very close to each other, the drag reduction caused by the slip, $DR\% = (\bar{\tau}_{\text{no-slip}} - \bar{\tau}_{\text{slip-defined}})/\bar{\tau}_{\text{no-slip}}$, is very small, where $\bar{\tau}_{\text{no-slip}}$ and $\bar{\tau}_{\text{slip-defined}}$ are the average shear stress at the air-water interface with no-slip and slip-defined boundary conditions, respectively. When the air-water interface was concave, $\triangle d < 0$, the drag reduction generated by the slip was quite small, dropping below 0.4% as shown in Fig. 6b. It is because the value of positive slip velocity at the side area of the interface approximates the one of the reverse slip at the center area, as mentioned above. So the average slip velocity is very close to zero and the effect of the slip is very weak. When the air-water interface was convex, $\triangle d > 0$, the drag reduction was found to increase with the interface deflection increasing, because the positive slip at the center area becomes larger than the negative slip at the side area, as shown in Fig. 5c. However, the maximum drag reduction was still predicted to be only 1.2% when it is convex.

As mentioned above, the slip length at the air-water interface is finite because of the limited slip velocity as observed in our experiment. To calculate the experimental slip length at the air-water interface, the shear stress at the interface obtained by the simulation shown in Fig. 6a was required. The local slip length at the interface equals:

$$b_l(y) = \mu u_s(y)/\tau(y), \tag{1}$$

where $y$ is the spanwise position at the interface as shown in Fig. 1b. The average slip length of the entire air-water interface equals:

$$\bar{b}_a = \frac{\int_0^{w_g} b_l(y)\frac{du}{dz}(y,0)dy}{\int_0^{w_g} \frac{du}{dz}(y,0)dy} = \frac{\int_0^{w_g} \frac{\mu u_s(y,0)}{\tau(y,0)}\frac{du}{dz}(y,0)dy}{\frac{1}{\mu}\int_0^{w_g} \tau(y,0)dy} = \frac{\mu \int_0^{w_g} u_s(y)dy}{\int_0^{w_g} \tau(y,0)dy}. \tag{2}$$

The local slip length at the center of the air-water interface $b_l(0.5W_g)$ and the average slip length across the air-water interface $\bar{b}_a$ are plotted in Fig. 6c. Similar to the trends observed with the slip velocity, both the local and average slip lengths were found to increase with increasing the deflection of the air-water interface. Note that the value of the average slip length was much smaller than expected, i.e., not only finite buy also roughly two orders of magnitude smaller than the interface width. At concave interfaces, $b_l$ (0.5$W_g$) was negative due to the reversed flow at the center of the air-water interface. Note that with increased negative surface deflection, $\triangle d < 0$, the average slip length approaches zero, indicating a loss of drag reduction for concave interfaces which is consistent with the results in Fig. 6b. For positive values of interface deflection beyond, $\triangle d \geq 30$ μm, the average slip length was found to be significantly smaller than slip length along the center of the air-water interface in part because of the presence of the reversed flow near the side walls of the air-water interface.

As mentioned earlier, the presences of a negative surface tension gradient along the streamwise direction would possibly cause the reduction of the slip at the air-water interface. To test the validity of this hypothesis, another simulation was performed where the interfacial boundary condition on the slip velocity was replaced with a uniform shear stress ($\bar{\tau}_{AW}$) standing for a constant surface tension gradient ($\partial \gamma / \partial x$), and the resulting velocity profiles were compared to the experimental measurements. This boundary condition was named shear-defined. $\bar{\tau}_{AW}$ was selected using the shear stress at the interface obtained by the simulation shown in Fig. 6a. The interfacial tension along the spanwise direction at the interface was treated as constant in the simulations, $\partial \gamma / \partial y = 0$. It is reasonable as the spanwise component of the slip velocity at the interface is much smaller than the streamwise one as shown in Fig. 5 and Video3.mp4 and Video4.mp4 of the supporting information [38].

The slip velocity profile resulting from the implementation of a constant shear stress across the interface is shown in Fig. 6d. It agrees well with the experimentally observed slip velocity profiles shown in Fig. 5c. The applied surface tension gradient (or shear stress) reduced the slip velocity significantly with a maximum slip velocity of $u_s = 0.06U_\infty$ observed for convex interfaces with $\triangle d > 0$ and $u_s = -0.054U_\infty$ for concave interfaces with $\triangle d < 0$. It should be noted that, the backward slip emerges in the simulation and, similar to the experiments, it appears along the area near the side walls of the interface for the convex air-water interfaces and along the center of the concave air-water interfaces. As shown in the simulations of Fig. 6d, the location of the reversed flow changes with the interface deflection, because the shear stress along the center of the air-water interfaces imposed by the flow decreases as it deflected into the groove as shown in Fig. 6a. By deflecting the interface downward from 124 μm to -56 μm, the local shear stress difference between the center line ($\tau(0.5W_g)$) and the average one ($\bar{\tau}_{AW}$), $(\tau(0.5W_g) - \bar{\tau}_{AW})/\bar{\tau}_{AW}$, changes from 5.7% to -2.3 %. Eventually, the strength of the Marangoni flow surpasses the shear stress induced by the flow along the centerline of the air-water interface and the flow reverses there. It is worth noting that, the interplay between the slip flow and the surface tension gradient within the air-water interface was neglected in this simulation. This simplification causes the singularity at the edge of the air-water interface and a larger value of negative slip velocity than the real one.

The origin of the surface tension gradient is still unclear. One of the possible causes may be the particle concentration variation along the air-water interface [47-51], which has been investigated here. The slip velocity at the interfaces with different concentrations has been performed. As shown in Fig. 7, the slip velocity at the air-water interface is still extremely small, $u_s(0.5y/W_g) < 1.5\% U_\infty$, even we decrease the number of particles to ~75 per square millimeter at the interface. What's more, the maximum of the slip velocity has been shown to be independent of the particle concentration as shown in Fig. 7d. The slip velocity at the interface with particle concentration as high as 1000 per square millimeter is very close to the one as sparse as 75 per square millimeter, as shown in Fig. 7c and Video5.mp4 and Video6.mp4 of the supporting information [38]. This experimental result implies that the interfacial tension gradient along the interface does not come from the particle concentration variation within the air-water interface.

One of the other causes of the surface tension gradient may come from the unknown surface-active agents in the channel or the solution of the tracer particles. The variation of the interfacial tension from the beginning to the end of the channel can be estimated from the numerical simulations to be only $\Delta\gamma = \bar{\tau}_{AW} \cdot L_g = 53.5$ μN/m. It suggests that such a small surface tension gradient along the air-water interface is sufficient to nearly eliminate all drag reduction and slip. It is consistent with the very recent work done by Peaudecerf et al. [27]. They demonstrated that the maximum Marangoni number at an air-water interface, $Ma = 2RT\Gamma / \mu U_\infty$, could reach as high as from $10^3$ to $10^6$ even for a clean micro channel, where $R$, $T$, $\Gamma$ are the gas constant, temperature, interfacial concentration of the surfactant at the interface, respectively. They showed that a force balance at the interface could be always observed: $L_g(1-u_s/U_\infty)/H \sim Ma$, even if the bulk concentration of surfactant being smaller than ~ $10^{-9}$ Mol/m$^3$. That low value of the concentration of surfactant cannot be avoided even by a standard microfluidic cleaning procedures.

Even though the above experiments show that the slip velocity or slip length is significantly smaller than the main velocity or surface-micro-structure scale, a great deal of previous work has shown a much larger drag reduction on the superhydrophobic surfaces and the existence of a large slip velocity at the air-water interface [2,3,7,52]. This inconsistency, which we believe results from the presence of surface tension gradients at the air-water interface, was further investigated by modifying the channel geometry to eliminate the presence of the head wall and create continuous air-water interface in a flow with closed streamlines. This was done using the rotating channel flow geometry shown in Fig. 1c. When the fluid flows along the continuous annular air-water interface without any interruptions or stagnation points in the flow direction, particle accumulations along the interface are not expected nor are surface tension gradients. As a result, in this geometry a large slip velocity, comparable to the main flow velocity was expected. As shown in Figs. 8a and 8b (see also Video7.mp4 of the Supporting Information [38]), the maximum of the slip velocity reaches as high as $u_s = 0.19 U_\infty$, where $U_\infty = \Omega r_c$, is the average of the angular velocity of the top rotating plate, with definitions of $\Omega$ and $r_c$ schematically shown in Fig. 1c. It is important to note that no instances of reversed flow were observed for this geometry.

The analytical solution of the slip velocity profile at the shear free interface along the spanwise direction was derived by Philip [53,54], $u_s(y,0) = \frac{\tau_\infty}{\mu}\sqrt{\frac{1}{4}W_g^2 - y^2}$, where $\tau_\infty$, $W_g$, $\mu$ were the shear stress parallel to the plate at infinity, interface width and viscosity, respectively. $y$ is the spanwise position at the interface and $y = 0$ is the center of the interface. In the flow shown in Fig. 1c, the local slip velocity at the interface could be transformed to

$$u_s(r) = \frac{U_\infty(r)}{H}\sqrt{\frac{1}{4}W_g^2 - (r-r_a)^2}, \tag{3}$$

by assuming $\tau_\infty \approx \mu U_\infty(r)/H$, where $r_a$, $r$, $H$, and $U_\infty = \Omega r$, are the radius of the center line of the interface, local radius, the gap distance between the top and bottom plate, and the local tangential velocity of the top plate, respectively. As shown in Fig. 8b, the experimental slip velocity in experiments and the one solved numerically by simulations at the continuous interface agree very well with the analytical results shown in Eq. (3).

The average slip velocity across the interface is

$$\bar{u}_s = \frac{1}{W_g}\int_{r_c-W_g/2}^{r_c+W_g/2} u_s(r)\mathrm{d}r = \frac{\pi W_g \Omega r_c}{8H}. \tag{4}$$

As shown in Figs. 8c and 8d, all these experimental results are qualitatively similar to the predictions of the numerical and analytical results with an overprediction by 10 - 30%. The presence of the surfactants on the air-water interface still appears to have a modest impact on slip even in the continuous geometries. What's more, the air-water interface in this flow is concave which may lower the slip velocity slightly. By contrast, however, the slip velocities observed in the previous channel with the rectangular air-water interface are two orders of magnitude lower than the predictions of simulations. It is clear that, at the annular air-water interface, the lack of surfactant concentration variation has eliminated the surface tension gradient resulting in a nearly full slip expected along the superhydrophobic surface.

## V. CONCLUSIONS

In conclusion, the effect of the surface tension gradient (Marangoni stress) along a millimeter sized air-water interface has been investigated systematically. For long, rectangular air-water interfaces, the interfacial shear stress was found not only to reduce the slip velocity, but also give rise to a negative interfacial slip velocity moving opposite to the main flow direction. Both experiment and numerical simulation showed that the curvature of the interface had a significant impact on the velocity profile across the air-water interface. For convex interfaces, the reverse flow was observed to occur along the side walls of the air-water interface. While for a concave interface, the reverse flow was observed along the center of the air-water interface. The magnitude of the slip was found to be greatly reduced in these cases with the average slip length of the air-water interface found to be less than 0.4% of the interface width and the maximum drag reduction found only to be 1.2%. These values were more than an order of magnitude smaller than the predictions for a shear-free air-water interface and show the importance that the Marangoni flow can have on the effectiveness of superhydrophobic surfaces. By changing the interface geometry from discontinuous to continuous, it was shown that the effect of the Marangoni stress on the slip could be mitigated. At the annular air-water interface, the surface-active agent distribution was homogenous so that no surface tension gradient was induced and a large slip velocity was observed in good agreement with the predictions of the numerical simulation imposing a shear-free boundary condition on the air water interface. In this case, a slip velocity as large as 30% of the main flow was achieved. Thus, our experiments clearly demonstrate the presence of Marangoni stress at the air-water interface, which can have a major impact on the drag reduction capabilities of the superhydrophobic surfaces. This is especially true for surfaces with pockets of discrete, discontinuous interface where flow-induced concentration gradients can build up resulting in Marangoni flows that resist flow.

**ACKNOWLEDGEMENTS**

This research was supported by National Science Foundation of China (No. 51679203), Natural Science Basic Research Plan in Shenzhen City of China (No. JCYJ20160510140747996), the Natural Science Basic Research Plan in Shaanxi Province of China (No. 2016JM1002) and the Doctorate Foundation of Northwestern Polytechnical University (No. CX201206). The work has also been partially funded by the US NSF Grant (No. 1462499) and the Alexander von Humboldt Foundation. We would like to thank the reviewers of this paper, whose comments strengthen our analysis and make our results and conclusions more rigorous.

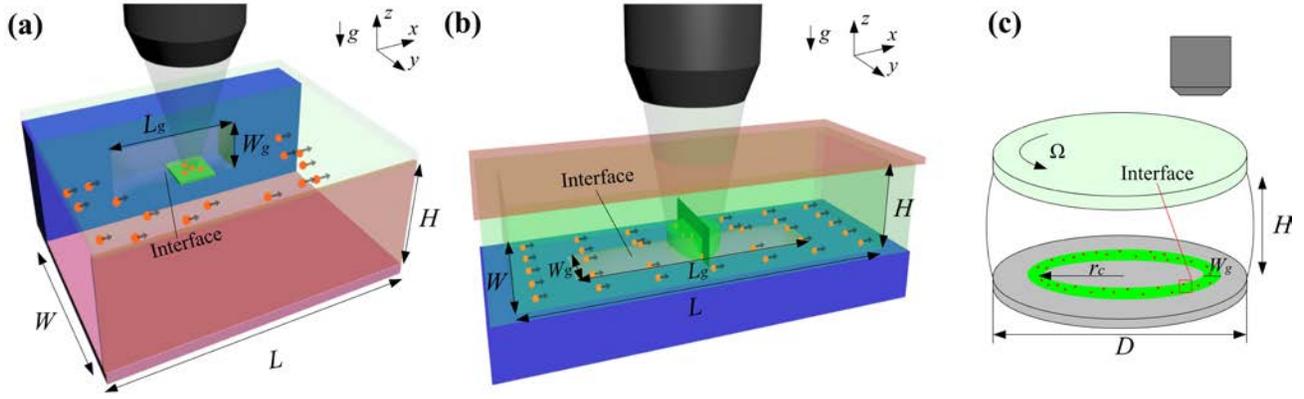

**FIG. 1**: Schematic diagram of the flow channels used in experiments. For the channel shown in (a), a horizontal groove was introduced at the center of the sidewall which was then covered by the vertical air-water interface after water is injected. The channel in (a) is 5.5 mm wide (*W*), 3.3 mm high (*H*), and 58 mm long (*L*), with the groove size being 1.0 mm wide ($W_g$), 5.0 mm long ($L_g$) and 5.0 mm deep ($D_g$). The channel in (b) contains a vertical groove centered at the bottom of the channel which is covered by a horizontal air-water interface. The channel in (b) is 5.0 mm wide (*W*), 2.0 mm high (*H*), and 56 mm long (*L*), with the groove size being 1.0 mm wide ($W_g$), 15.0 mm long (*Lg*) and 2.0 mm deep ($D_g$). The geometry in (c) is a torsional shear flow. The bottom plate in (c), with diameter *D* = 40 mm, contained an annular groove which supports an air-water interface, with radius of center line $r_c$ = 8 mm. The width of the annular groove ranges between 0.94 mm < $W_g$ < 1.83 mm. The top plate was rotated and the gap between the top and bottom plate ranges between 1.0 mm < *H* < 4.0 mm. The slices highlighted in green in (a) and (b) demonstrate the plane within the focal plane of the microscope.

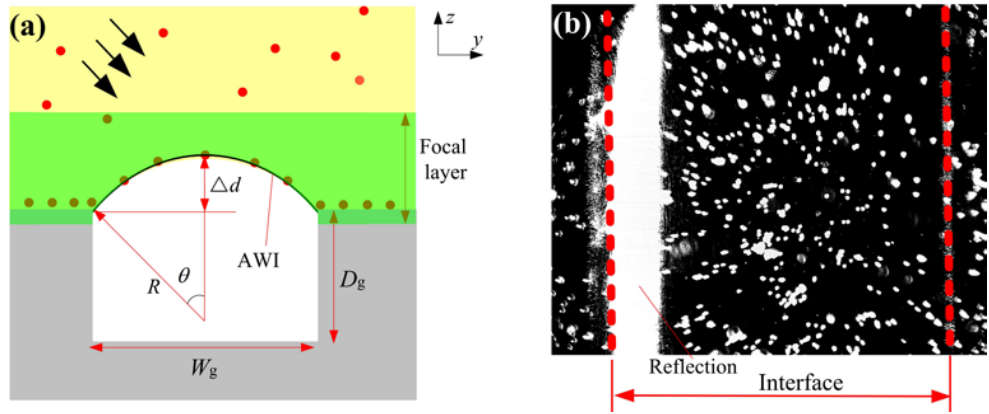

**FIG. 2.** (a) Schematic diagram of the cross-section of the channel showing the geometry of the air-water interface as it deflected out of the groove (convex). (b) Representational experimental image of the convex air-water interface showing the particle image velocimetry tracer particles and a reflection of the illuminating laser light sheet from the air-water interface.

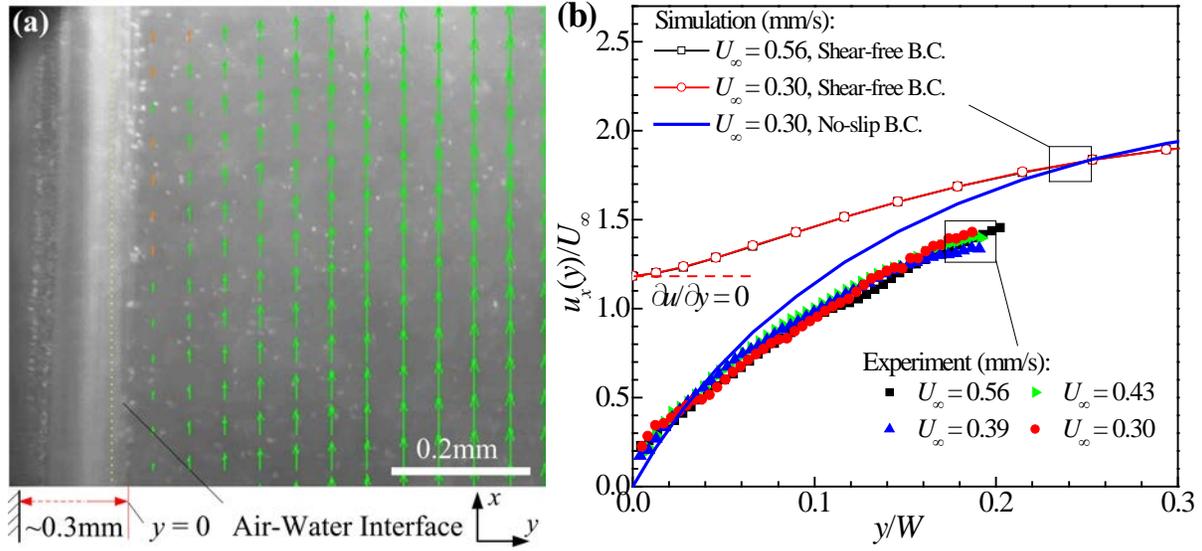

**FIG. 3.** Measurements of the velocity profile normal to the air-water interface. (a) An image of fluorescent particles near the air-water interface with the velocity vector field superimposed for the flow in the channel shown in Figure 1a with $U_\infty = 0.56$ mm/s. (b) The velocity profile along the line normal to the interface. The velocity was time-averaged over 500 frames. The solid symbols (■▶▲●) in (b) indicate the experimental data at different bulk velocities provided in the legend. The hollow symbols (—□—, —○—) in (b) represent the results of numerical simulations with a shear-free boundary condition imposed on the interface while the solid line (—) represent the simulation results for an interface with a no-slip boundary condition. The velocity and position have been normalized by bulk velocity ($U_\infty$) and channel width ($W$), respectively. In this experiments, $1.0 < Re < 1.8$ and $0.41 \times 10^{-5} < Ca < 0.76 \times 10^{-5}$.

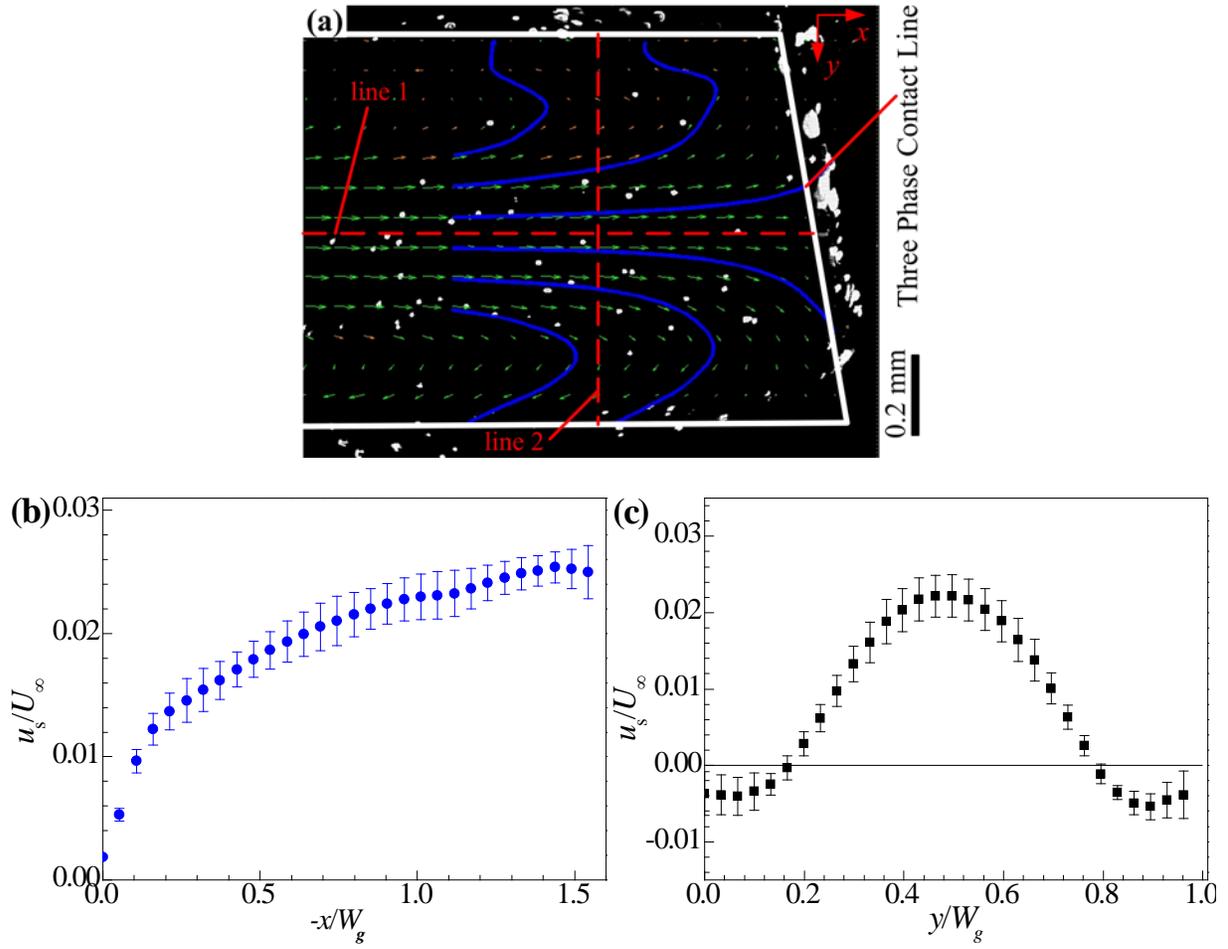

**FIG. 4.** Details of the interfacial velocity profiles for flow in the channel shown in Figure 1b for an interfacial deflection of $\triangle d$ = 92 μm and an average flow speed of $U_\infty$ = 1.06 mm/s. The velocity vector field in (a) just upstream of where the termination of air-water interface is shown in green with streamlines superimposed in blue. The normalized slip velocity along the center of the groove (along line 1) is shown in (b). The streamwise component ($x$) of the slip velocity across the air-water interface (along line 2) is shown in (c). The negative value in (c) means the slip velocity is opposite to the main flow direction. In this condition, $Re$ = 2.1 and $Ca$ = $1.3 \times 10^{-5}$.

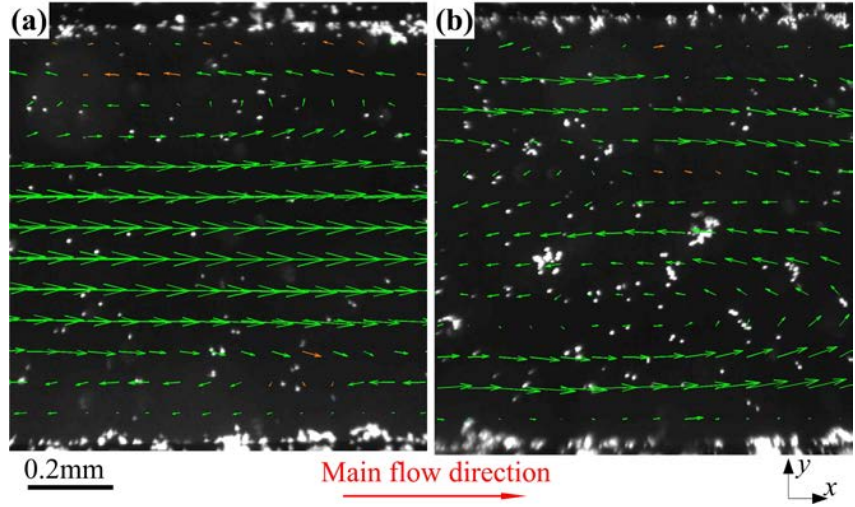

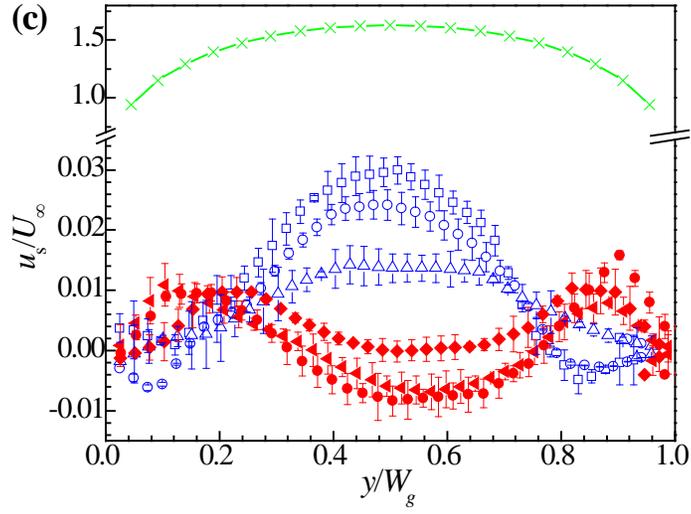

**FIG. 5.** Dependence of the slip velocity on the deflection of the air-water interface. The velocity vectors in (a) were at a convex interface with $\triangle d = 92$ μm and the ones in (b) were at a concave interface with $\triangle d = -56$ μm at the bulk velocity of $U_\infty = 1.06$ mm/s. The streamwise component ($x$) of the slip velocity along the crosswise direction ($y$) at the center of the air-water interface is shown in (c) for a series of interfaces with curvature. The data includes: □, ○, △, ◆, ◀ and ● correspond to $\triangle d =$ 124 μm, 92 μm, 60 μm, 0 μm, -28 μm and -56 μm, respectively, along with simulation result, —×—, based on the shear-free assumption at the air-water interface. In the experiments, $Re = 2.1$ and $Ca = 1.3 \times 10^{-5}$.

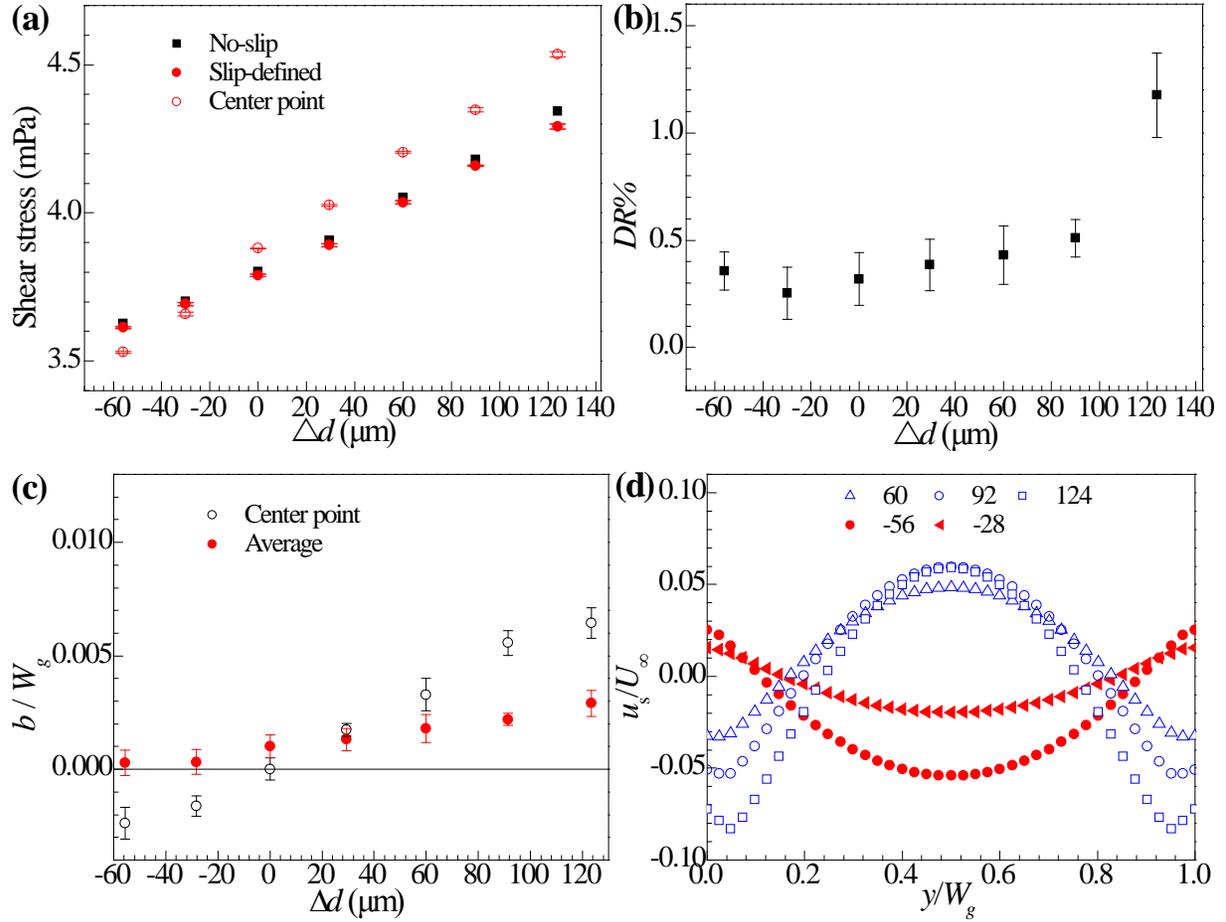

**FIG. 6.** Simulation results of the interfacial flow at the interfaces with different deflections. (a) is the overall shear stress. (b) is the overall drag reduction. (c) is the normalized slip length. And (d) is the normalized local slip velocity. For simulations in (a), (b) and (c), the experimental slip velocity shown in Figure 5c was specified at the air-water interface as the user-defined slip boundary condition. The solid squares (■) and solid cycles (●) in (a) respectively stand for the average shear stress at interface with the no-slip boundary condition and user-defined slip boundary condition, while the hollow cycles (○) sand for the local shear stress at the center of the interface with the user-defined slip boundary condition. Drag reduction in (b) was calculated based on the average shear stress at the interfaces between the no-slip (■) and slip-defined (●) data in (a). The solid cycles (●) and hollow cycles (○) in (c) denote the average slip length across the entire air-water interface and the local slip length at the center of interface. For simulations shown in (d), a shear stress was specified at the air-water interface as the boundary condition and the value of the stress was calculated from the numerical value shown in (a). More specifically, the shear stress applied in (d) was 3.61, 3.69, 3.79, 4.04, 4.16, 4.29 mPa, for interfaces with deflection of -56, -28, 0, 60, 92, 124 µm, respectively.

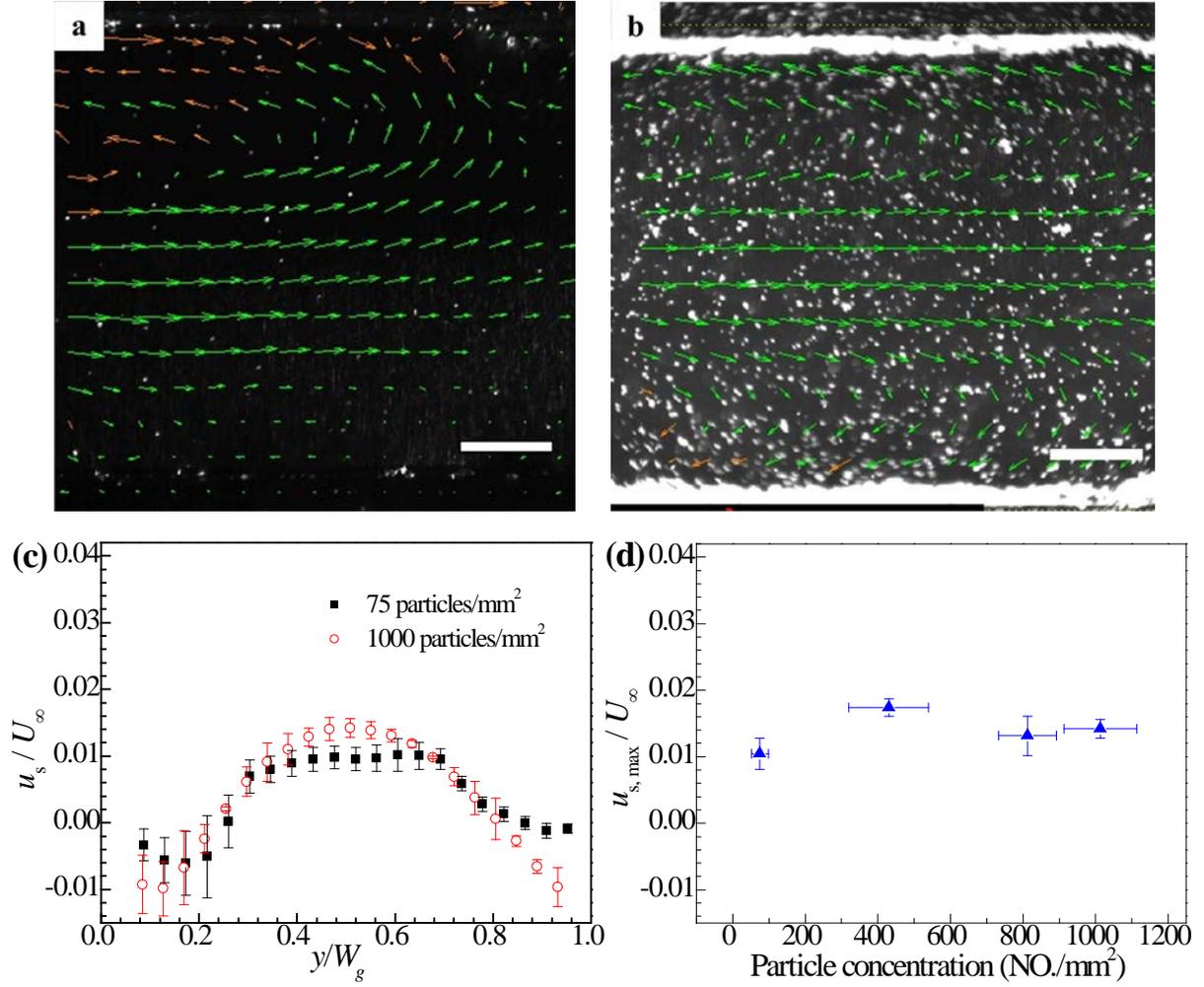

**FIG. 7.** Interfacial velocity vector at the air-water interfaces with different concentration of particles. The number of particles in (a) and (b) is 75 mm$^{-2}$ and 1000 mm$^{-2}$, respectively. The streamwise component of the slip velocity along the spanwise direction in (a) and (b) is shown in (c). The maximum of the slip velocity changes as a function of the particle concentration at the air-water interface is shown in (d). The interface was kept constant with convex deflection $\triangle d$ = 60 μm in (d). Scale bar in (a) and (b) is 0.2mm.

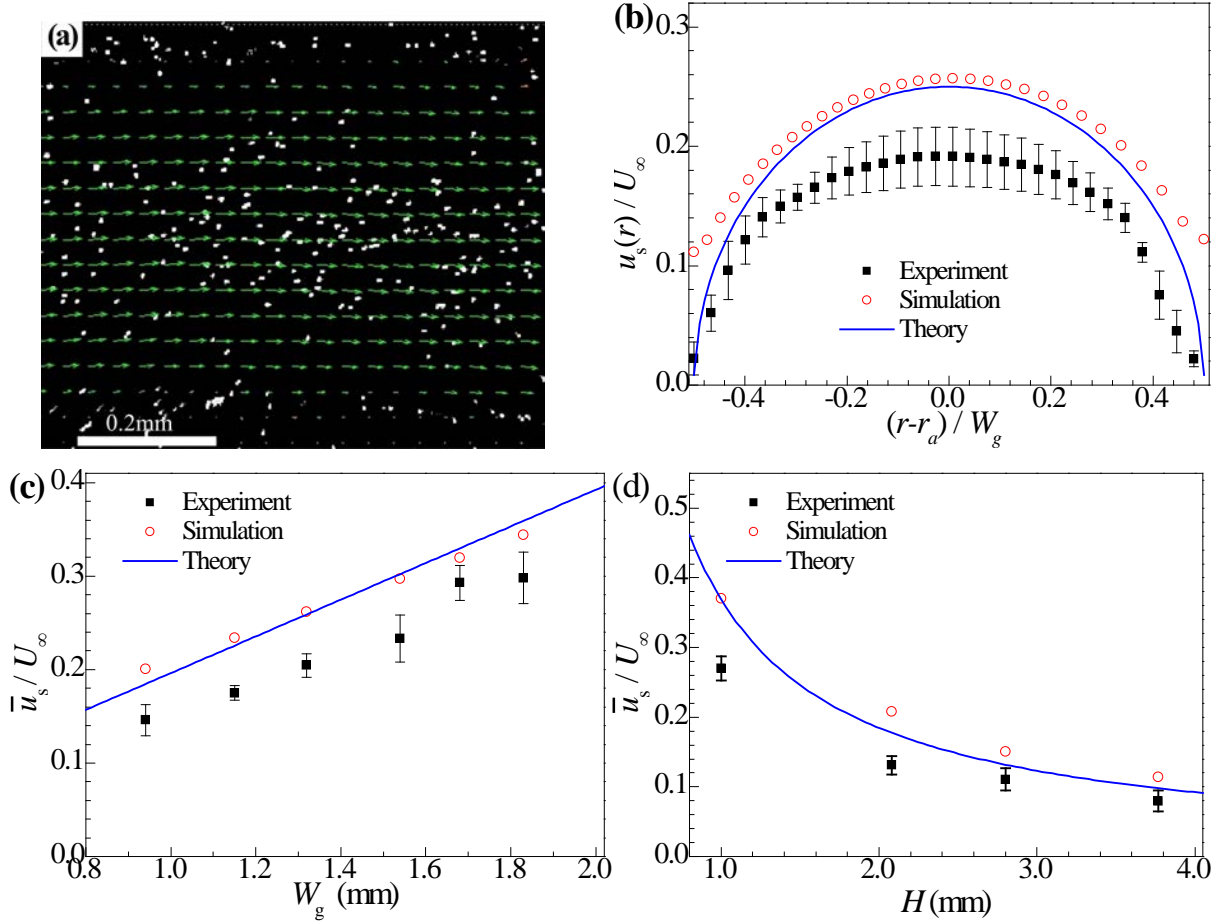

**FIG. 8.** Measurements of slip velocity along the annular air-water interface in the torsional shear flow shown in Figure 1c. The vector field of slip velocity across the air-water interface is shown in (a). The azimuthal slip velocity is plotted as a function of radial position across the air-water interface in (b). The average of the normalized slip velocity at the air-water interface is plotted as a function of the interface width and channel height in (c) and (d), respectively. The channel height in (c) was kept constant as $H = 2.0$ mm, while the interface width was kept constant as $W_g = 0.94$ mm in (d). The solid squares in (b), (c) and (d), are experimental data while the hollow cycles correspond to simulation results. The solid line in (b) is the prediction of Eq. (3). The solid lines in (c) and (d) are predictions of Eq. (4) for the average slip velocity.

28